\documentclass[pdflatex,sn-mathphys-num]{sn-jnl}

\usepackage{graphicx}%
\usepackage{multirow}%
\usepackage{amsmath,amssymb,amsfonts}%
\usepackage{amsthm}%
\usepackage{mathrsfs}%
\usepackage[title]{appendix}%
\usepackage{xcolor}%
\usepackage{textcomp}%
\usepackage{manyfoot}%
\usepackage{booktabs}%
\usepackage{algorithm}%
\usepackage{algorithmicx}%
\usepackage{algpseudocode}%
\usepackage{listings}%

\theoremstyle{thmstyleone}%

\theoremstyle{thmstyletwo}%

\theoremstyle{thmstylethree}%

\begin{document}

\title[Hash Chemistry: Minimal Models]{Hash Chemistry: Minimal Models for Evolutionary Growth of Complexity}

\author*[1]{\fnm{Ilya} \sur{Horiguchi}}\email{ilya-horiguchi@g.ecc.u-tokyo.ac.jp}

\author*[2,3,4]{\fnm{Hiroki} \sur{Sayama}}\email{sayama@binghamton.edu}

\affil*[1]{\orgname{Graduate School of Arts and Sciences, The University of Tokyo},
\orgaddress{\street{3-8-1 Komaba}, \city{Meguro-ku}, \state{Tokyo}, \postcode{153-8902}, \country{Japan}}}

\affil[2]{\orgname{Binghamton Center of Complex Systems, Binghamton University, State University of New York},
\orgaddress{\city{Binghamton}, \state{NY}, \postcode{13902}, \country{USA}}}

\affil[3]{\orgname{Systems Engineering, Cornell University},
\orgaddress{\city{Ithaca}, \state{NY}, \postcode{14850}, \country{USA}}}

\affil[4]{\orgname{Waseda Innovation Lab, Waseda University},
\orgaddress{\city{Tokyo}, \country{Japan}}}

\abstract{
Hash Chemistry is a family of minimalistic evolutionary models in which a deterministic hash function assigns a scalar score to entities of arbitrary size, opening a combinatorially vast possibility space (a ``cardinality leap''). Since its introduction, the idea has been realized in several settings, from the original spatial formulation to a fast non-spatial variant and then to structural cellular models. Here we review the Hash Chemistry family as a coherent modeling framework and use it to explore how minimal systems can demonstrate the mechanisms behind multiscale open-ended evolutionary dynamics. The most recent model, Structural Cellular Hash Chemistry (SCHC), successfully demonstrated multiscale ecological interaction/adaptation and complexity growth of replicators in a computationally efficient manner. In this study, we first extend SCHC to incorporate spatial locality and dyadicity of competitive interactions among replicating structures. We show this extension substantially enhances SCHC's evolutionary dynamics. Furthermore, we explore SCHC in a significantly larger spatial domain using a GPU-accelerated implementation. We show that the size of the space acts as a control parameter for a stochastic, nucleation-like transition between a compact-replicator regime and a runaway size-dominance regime, and we separate the responsible mechanism into a non-spatial, size-biased sampling feedback and a finite-size spatial effect. Altogether, these results illustrate the rich potential of Hash Chemistry as a minimal, mechanistically transparent testbed for studying open-ended evolution across scales.
}


\maketitle

\section{Introduction}\label{sec:intro}

The open-endedness of artificial evolutionary systems has been one of the most actively discussed agendas in the artificial life research community for the last decade \citep{taylor2016open,banzhaf2016defining,stanley2019open,packard2019overview,stepney2021modelling,stepney2023open,borg2023evolved}. Recently, there was a surge of interest in developing minimalistic distributed artificial chemistry \citep{dittrich2001artificial,banzhaf2015artificial} models for the exploration of open-ended evolutionary dynamics \citep{chan2023,nichele2024special,sayama2024review,yang2024emergence,rodriguez2024saucerful}. The emergent, self-organizing nature of such distributed artificial chemistry models has significant potential for the open-endedness research since they often have fewer constraints or built-in mechanisms than other more standard evolutionary or learning algorithms. 

We have contributed to this growing body of artificial chemistry-based open-endedness literature through a series of ``Hash Chemistry'' models \citep{sayama_hashchem_2019,sayama_nonspatial_2024,sayama_cellular_2024,sayama_schc_2025,sayama_spatialdyadic_2026,horiguchi_alife2026}. Hash Chemistry is a family of minimalistic evolutionary systems in which a deterministic hash function assigns a scalar score to replicating entities of arbitrary size, enabling a vast possibility space via combinatorial growth (``cardinality leap'') \cite{sayama_hashchem_2019}. A distinctive feature of Hash Chemistry is that the hash-derived score is deliberately \emph{not} identical to realized (ecological) fitness: it supplies an unbounded, structure-dependent scalar landscape, while what actually persists and reproduces is an emergent property of the interaction rules and the medium in which they act.

This paper has two aims. First, we review the Hash Chemistry family as a coherent modeling framework, tracing how the same core idea has been instantiated across progressively richer settings, including particle-based \citep{sayama_hashchem_2019}, non-spatial \citep{sayama_nonspatial_2024}, and most recently structural cellular \citep{sayama_schc_2025} versions, and what each instantiation reveals about open-ended and multiscale evolutionary dynamics. Second, we extend the most recent Structural Cellular Hash Chemistry (SCHC) to explore the effects of spatial locality and dyadicity of replicator interactions and the size of the space on the multiscale evolutionary dynamics of the system. We will then discuss implications for biological evolution and future research directions.

\section{The Hash Chemistry family of models}\label{sec:family}
There are several variations in the Hash Chemistry model family, but they share a common core: entities of arbitrary size are scored by a deterministic hash function, and higher-scoring entities are favored in a competition-and-replacement dynamic. They differ in whether and how the entities are embedded in space, and in what constitutes a replicating entity. The unifying idea is the \emph{cardinality leap} \cite{sayama_hashchem_2019}: by allowing higher-order entities (sets, multisets, or connected patterns) to form, the cardinality of the possibility space grows combinatorially with entity size, and a generic hash function serves as a universal ``oracle'' that can assign a scalar score to an entity of any size or scale. The family has evolved through a sequence of models that progressively add the properties a minimal open-ended evolutionary system is expected to display: continuous adaptation, unbounded growth of complexity (size of replicating patterns), nontrivial spatial ecological interactions, population diversity, and computational efficiency, ideally all at once. Table~\ref{tab:family} summarizes the progression.

\begin{table}[t]
\caption{The Hash Chemistry family of models. Each model is a step toward a minimal open-ended evolutionary system that simultaneously exhibits continuous adaptation, unbounded complexity (size) growth, nontrivial spatial ecological interactions, population diversity, and computational efficiency. Attributes follow the comparison in \cite{sayama_schc_2025}.}\label{tab:family}
\small
\begin{tabular}{@{}p{0.185\textwidth}c p{0.215\textwidth} p{0.34\textwidth}@{}}
\toprule
Model & Ref. & Representation & Status \\
\midrule
Original Hash Chemistry & \cite{sayama_hashchem_2019} & Spatial, groups of particles & Established the cardinality leap; computationally expensive; bounded complexity growth \\
Non-spatial Hash Chemistry & \cite{sayama_nonspatial_2024} & Well-mixed multisets & Fast; continuous adaptation and unbounded growth; no spatial interactions; low diversity \\
Cellular Hash Chemistry (prototype) & \cite{sayama_cellular_2024} & 2D grid, no explicit identity & Spatial interactions and low computational cost; but no meaningful adaptation or growth \\
Structural Cellular Hash Chemistry (SCHC) & \cite{sayama_schc_2025} & Connected components on a 2D grid & Achieves all desired properties at once; basis of the present study \\
Spatial-dyadic SCHC & \cite{sayama_spatialdyadic_2026} & SCHC $+$ local, dyadic interactions & Locality promotes growth; dyadicity promotes diversity \\
\botrule
\end{tabular}
\end{table}

\subsection{Original Hash Chemistry and the cardinality leap}\label{subsec:original}
\begin{figure}
\centering
\includegraphics[width=0.8\textwidth]{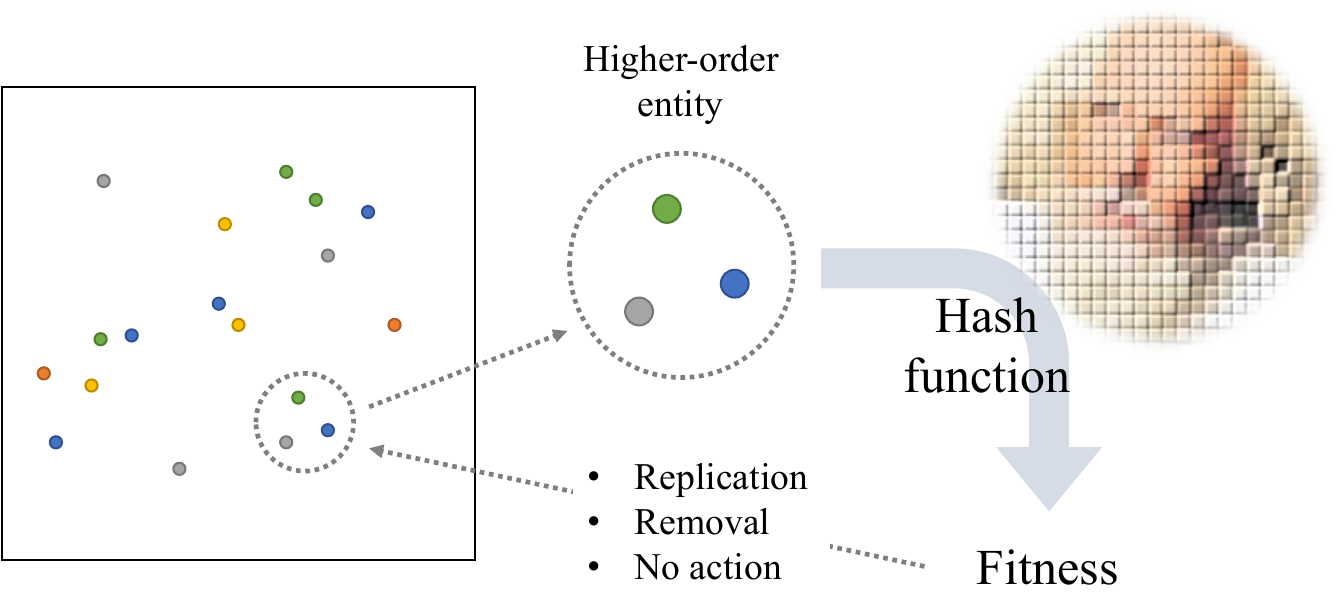}\\~\\
\includegraphics[width=0.24\textwidth]{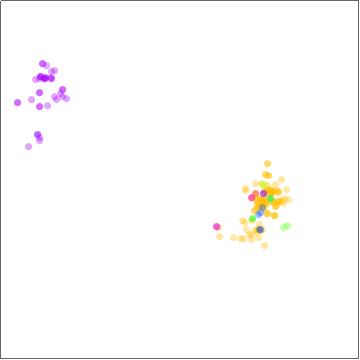}
\includegraphics[width=0.24\textwidth]{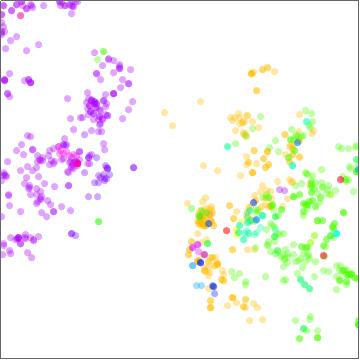}
\includegraphics[width=0.24\textwidth]{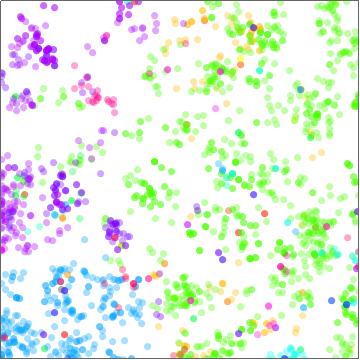}
\includegraphics[width=0.24\textwidth]{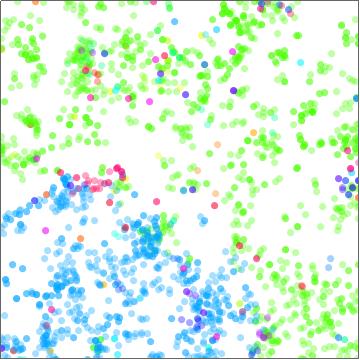}
\caption{Top: Schematic illustration of the original Hash Chemistry model. Bottom: Sample run of the Hash Chemistry model (from \cite{sayama_hashchem_2019}, with modification). Each frame shows a snapshot of the system at a certain time point (from left to right: $t = 30, 100, 300, 1000$). Each individual particle is represented as a dot in the space, with a color showing its type.}
\label{fig:original}
\end{figure}

The original Hash Chemistry \cite{sayama_hashchem_2019} introduced the cardinality-leap concept and demonstrated its effect on open-ended evolution. It is a spatial artificial chemistry with a finite continuous 2D space in which particles with finite discrete types are distributed and replicating. A hash function is used as a convenient oracle to evaluate the ``fitness'' of a group of particles of any arbitrary number that exist in spatial proximity, so that ever-larger composite entities can always have a chance to find higher-scoring variants, sustaining continued exploration and adaptation. Based on the hash score, the group of the particles can be replicated, removed, or stay intact (Fig.\ \ref{fig:original}, top). Repeated application of this simple extraction-replacement operation generates complex multiscale evolutionary dynamics (Fig.\ \ref{fig:original}, bottom). This model established the core idea of cardinality leap, but it was computationally very expensive because the local density of particles would rapidly become very large, causing enormous computational complexity for neighbor detection. It was also somewhat bounded in the complexity growth of replicating entities.

\subsection{Non-spatial Hash Chemistry}\label{subsec:nonspatial}
\begin{figure}
  \centering
\includegraphics[width=\textwidth]{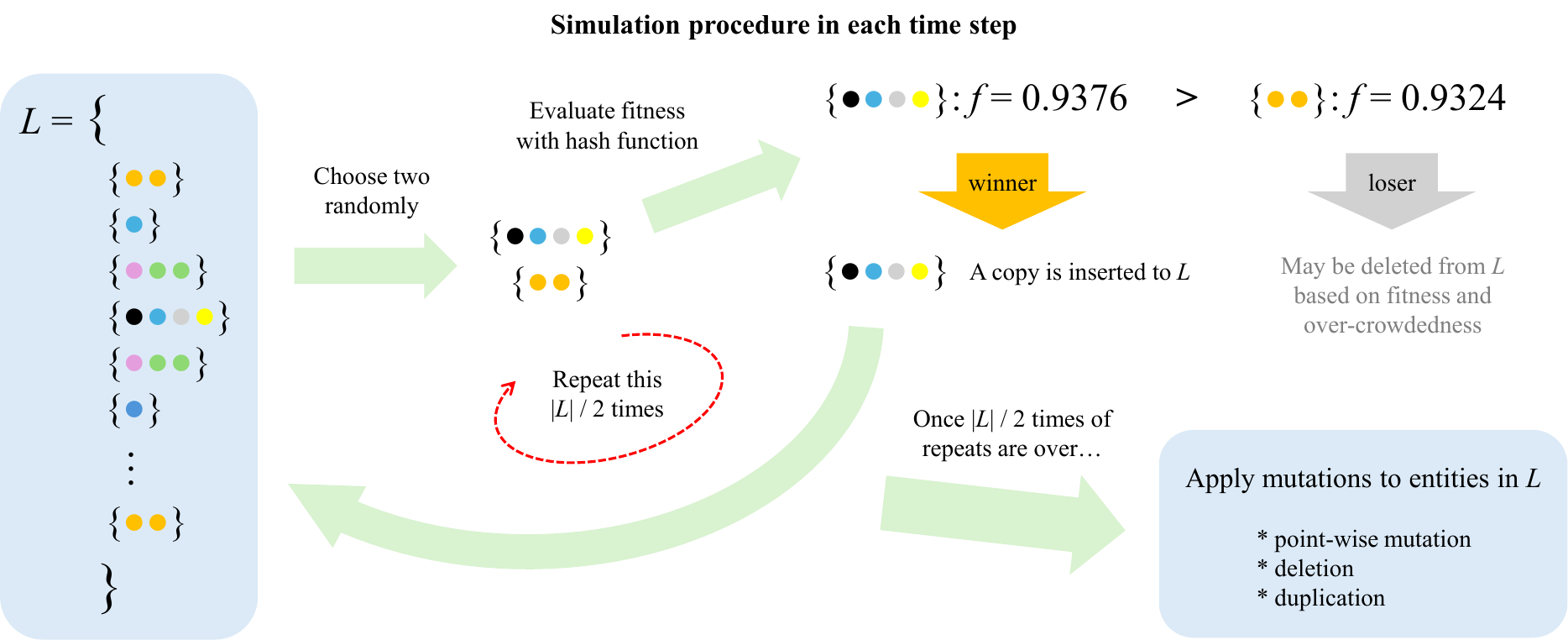}
\caption{Schematic illustration of the non-spatial Hash Chemistry model (from \cite{sayama_nonspatial_2024}, with modification).}
\label{fig:nonspatial}
\end{figure}

The non-spatial variant of Hash Chemistry \cite{sayama_nonspatial_2024} recasts replicating entities explicitly as multisets of particles rather than groups of spatially nearby particles. The population of replicators is represented as a list of multisets, and the evolution is simulated through repeated pairwise competition between two randomly selected multisets (Fig.\ \ref{fig:nonspatial}). This simple non-spatial setup gave a large speed-up and, for the first time in the family, a clear demonstration of continuous adaptation together with unbounded growth of complexity (size) of replicating entities. The simplification came at a cost, however: because the interactions are always only competitions between two entities randomly selected from the entire population and selection acts directly on individual hash values, the model does not exhibit any nontrivial spatial ecological interactions, and its overly strong selection pressure erodes population diversity. This model showed that space is not a necessary factor for open-ended evolution but removing it from the model makes the resulting evolutionary dynamics severely constrained and simplistic.

\subsection{Cellular Hash Chemistry (prototype)}\label{subsec:cellular}
The Cellular Hash Chemistry \cite{sayama_cellular_2024} is an unsuccessful prototype of a cellular variant which aimed to move the model in the complementary direction, placing replicating entities on a two-dimensional cellular grid. This restored multiscale spatial ecological interactions among evolving patterns and, with simplified four-way competition rules, was computationally very cheap. However, in this prototype, the entities lacked explicit individual identity, and the model failed to show meaningful adaptive evolution or complexity growth, leaving open whether a single minimal model could achieve spatial interactions and unbounded adaptation simultaneously. This nonetheless served as an important stepping stone for developing the next structural variant.

\subsection{Structural Cellular Hash Chemistry}\label{subsec:schc_model}
\begin{figure}
\centering
\includegraphics[width=\textwidth]{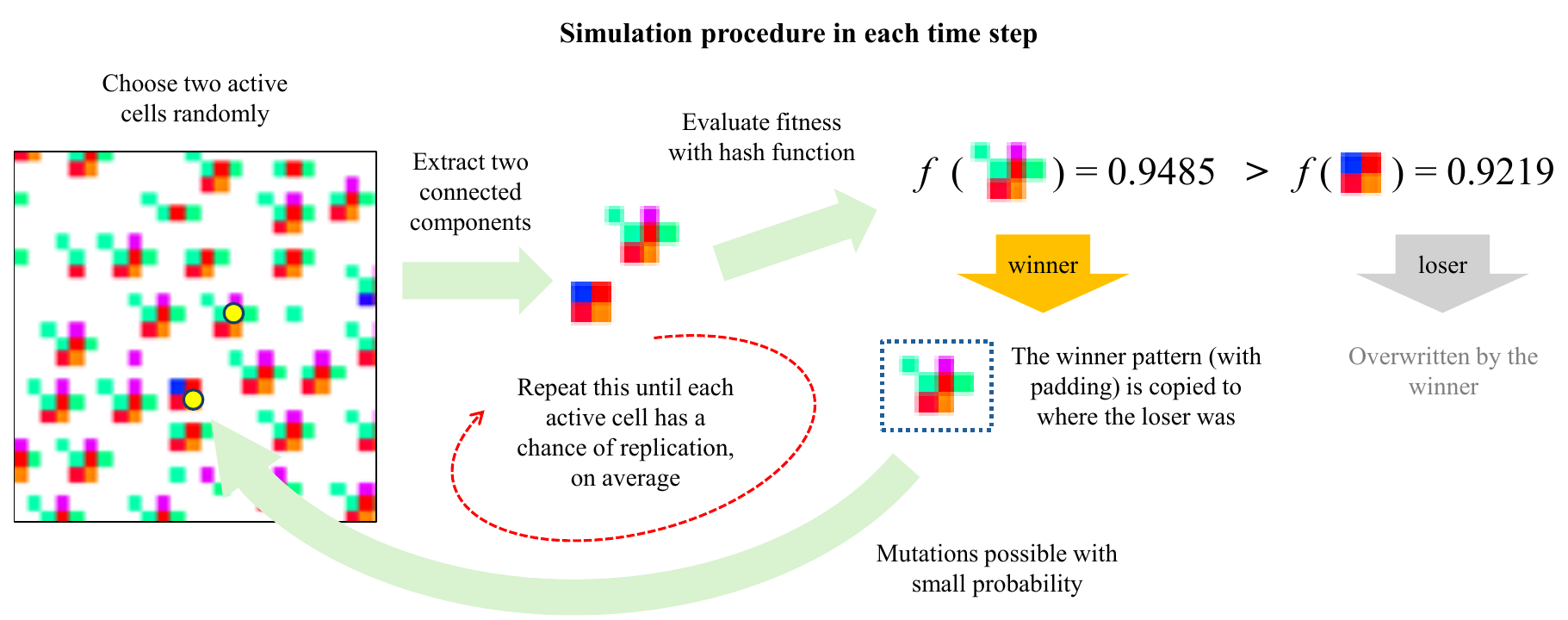}\\~\\
\fbox{\includegraphics[width=0.225\textwidth]{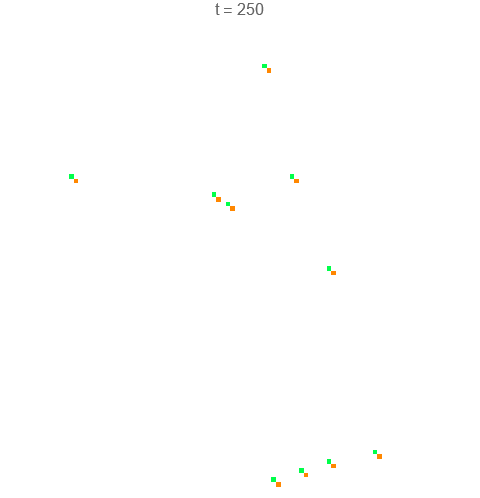}}
\fbox{\includegraphics[width=0.225\textwidth]{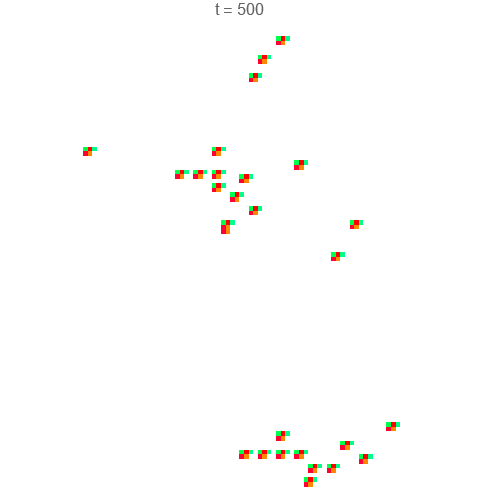}}
\fbox{\includegraphics[width=0.225\textwidth]{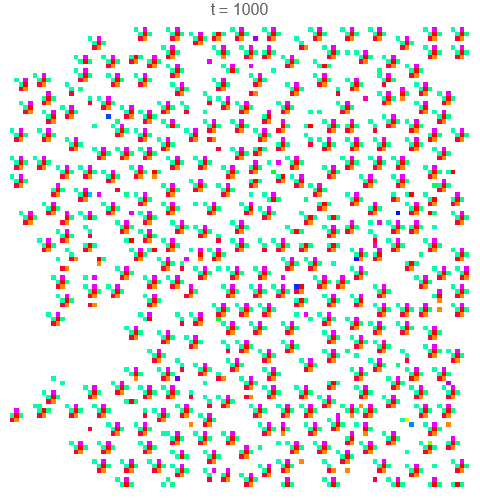}}
\fbox{\includegraphics[width=0.225\textwidth]{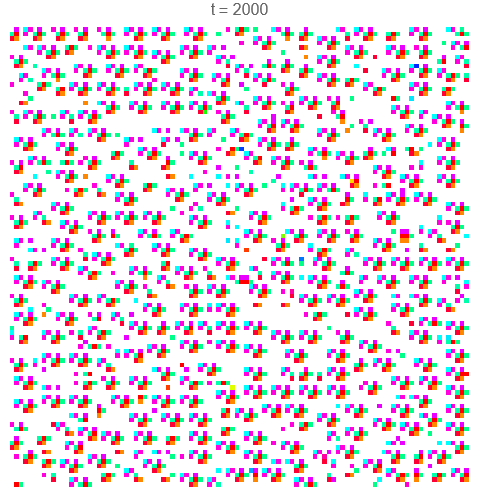}}
\caption{Top: Schematic illustration of the Structural Cellular Hash Chemistry (SCHC) model. Bottom: Sample run of SCHC (from \cite{sayama_schc_2025}, with modification). Each frame shows a snapshot of the system at a certain time point (from left to right: $t = 250, 500, 1000, 2000$). Colors represent different element types, and blank (white) spaces represent empty cells.}
\label{fig:schc}
\end{figure}

The Structural Cellular Hash Chemistry (SCHC) \cite{sayama_schc_2025} is the most recent variant of Hash Chemistry that successfully resolved the challenges the earlier versions faced. Its key device is to represent each replicating entity explicitly as a \emph{connected component} of the nearest-neighbor graph of active (non-empty) cells, i.e., cells linked to other active cells in their Moore neighborhood, on a $L \times L$ 2D grid. Evolution is simulated through pairwise competitions of two randomly selected components, as in the non-spatial version (Fig.\ \ref{fig:schc}, top). In each competition, two active cells are sampled uniformly at random from the whole grid, their components are identified and scored by a deterministic hash function, and the higher-scoring component replicates into the loser's location with per-cell death and mutation. 
The full update rules and our simulator implementations are given in the Methods (Section~\ref{sec:methods}).

SCHC achieved, for the first time in the family, all of the sought-after properties at once: continuous adaptation, unbounded complexity growth, nontrivial spatial ecological interactions, promoted diversity, and computational efficiency (Table~\ref{tab:family}; Fig.\ \ref{fig:schc}, bottom). A notable qualitative finding is that the population-average hash score initially rises but then declines once patterns form crowded clusters and begin to interfere spatially, signaling that realized fitness has departed from the hash score. SCHC is the main model the present study explores further in the following sections.

\section{Extension I: Introducing spatial locality and dyadicity of interactions}\label{sec:dyadic}

In the original SCHC, the competition between two replicating structures was determined solely by individually evaluated hash values, and the structures were randomly selected for competition from anywhere in the space. Ecological compatibility between different patterns or their spatial proximity did not matter, and therefore, the competition always took place globally and individually. These assumptions grossly ignored the possible importance of spatial locality and dyadicity of ecological interactions.

\begin{figure}
\centering
\includegraphics[width=\textwidth]{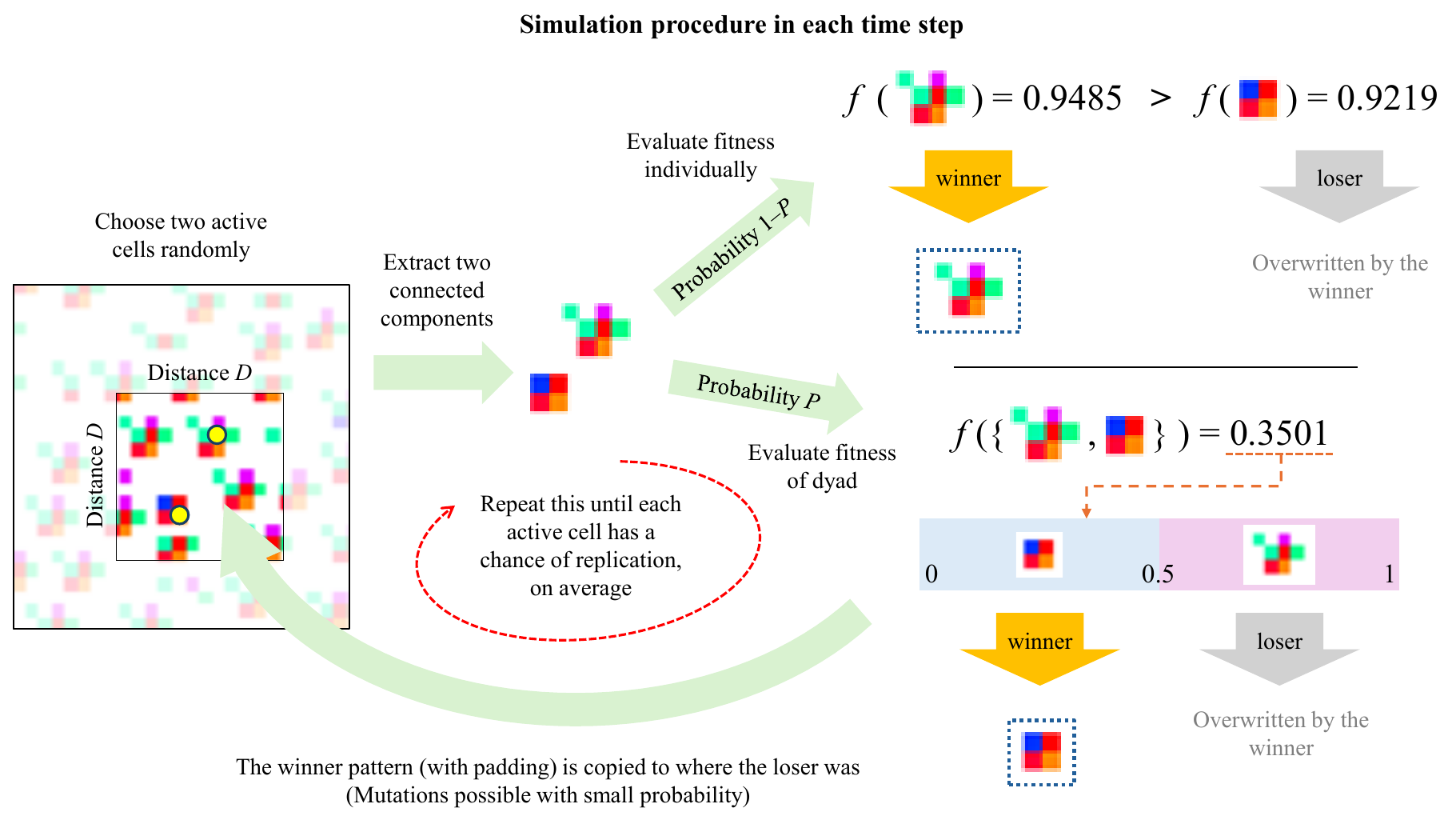}
\caption{Schematic illustration of the extended SCHC model with spatial and dyadic interactions.}
\label{fig:schc-SD}
\end{figure}

In this section, we extend SCHC by introducing the following two additional parameters that control the interactions between competing self-replicating patterns \cite{sayama_spatialdyadic_2026}:
\begin{itemize}
    \item \textbf{Spatial interaction range $D \in [1, L]$} (where $L$ is the linear size of the space): the maximal chessboard distance within which the competition between two patterns is allowed.
    \item \textbf{Dyadic interaction probability $P \in [0, 1]$}: the probability of dyadic hash evaluation when two patterns compete. 
\end{itemize}
The revised simulation algorithm that uses these two parameters is depicted in Fig.\ \ref{fig:schc-SD} and also described in more detail in Section \ref{sec:methods}. This parameterization contains the original model setting at one of the corners in the parameter space ($(D,P) = (L, 0)$) and enables exploration of different levels of spatial locality and dyadicity through systematic parameter sweep experiments. 

\subsection{Experiments}

We conducted a set of numerical experiments to investigate the effects of spatial locality of interactions and dyadic interaction probability on the evolutionary dynamics of SCHC. We used the same settings as in \cite{sayama_schc_2025}: $L = 100$ (linear size of space), $k = 1000$ (number of active cell types), and initial configurations that are almost empty but with 10 randomly generated initial active cells. We varied $P$ from 0 to 1 in increments of 0.2, and $D$ from $0.1L$ $(=10)$ to $L$ $(=100)$ in increments of $0.1L$ $(= 10)$. This makes the number of experimental conditions $6 \times 10 = 60$, and we conducted 30 independent simulation runs for each condition. The condition $(D, P) = (L, 0)$ corresponds to the original setting used in \citep{sayama_schc_2025}. Each simulation was run for 2,000 steps. 

To characterize the evolutionary dynamics observed in numerical simulations, we recorded the following three key measurements in this study:
\begin{enumerate}
\item[A.] Number of successful runs that showed meaningful population growth
\item[B.] Average number of active cells involved in each successful replication event
\item[C.] Cumulative number of unique replicating pattern types observed in successful replication
\end{enumerate}
Measurement A was characterized by the number of simulation runs (out of 30) that showed meaningful replication and population growth (as compared to extinction or absence of population growth), which characterized how robust self-replication and population growth was. Measurement B was calculated using the results from the last 500 steps of each simulation run, which captured the spontaneous size growth of replicating patterns occurring during the simulation. Finally, Measurement C was recorded at the end of the simulation, which quantified how much open-ended evolutionary exploration took place throughout the simulation. Data from simulation runs in which the population became extinct prematurely, did not show any meaningful growth, or accidentally exhibited explosive nucleation transition (there was just one such anomaly case for $(D, P) = (10, 0.6)$; such behaviors will be discussed in Section \ref{sec:schc}) were filtered out from the following analysis.

\subsection{Results}

\begin{figure}
\centering
\begin{tabular}{ll}
{\bf (a)} & {\bf (b)}\\
\includegraphics[height=1.4in]{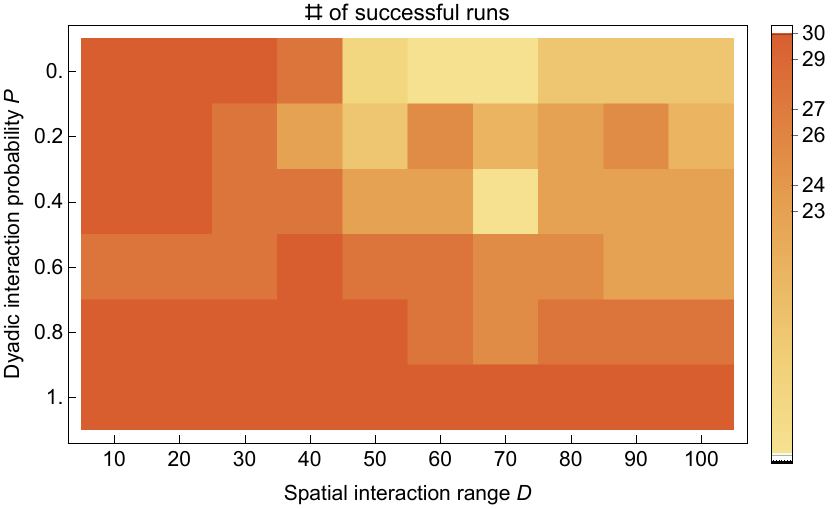} &
\includegraphics[height=1.4in]{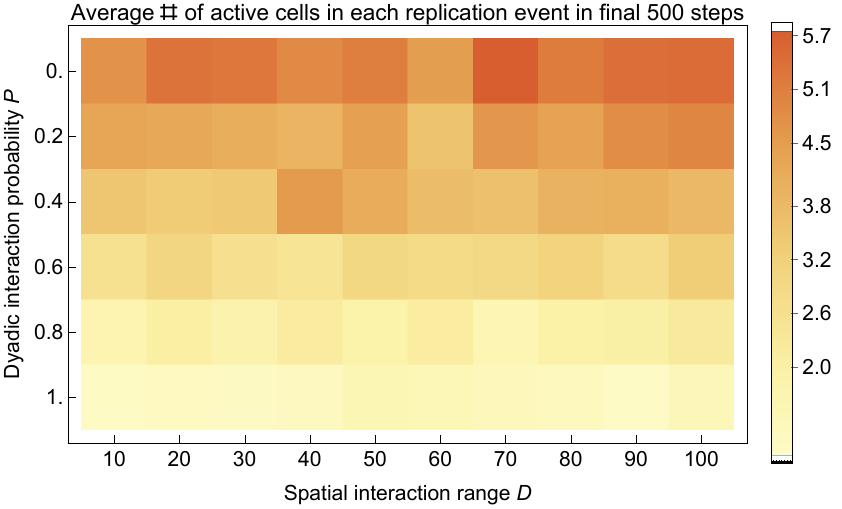}\\
~\\
{\bf (c)} & {\bf (d)}\\
\includegraphics[height=1.4in]{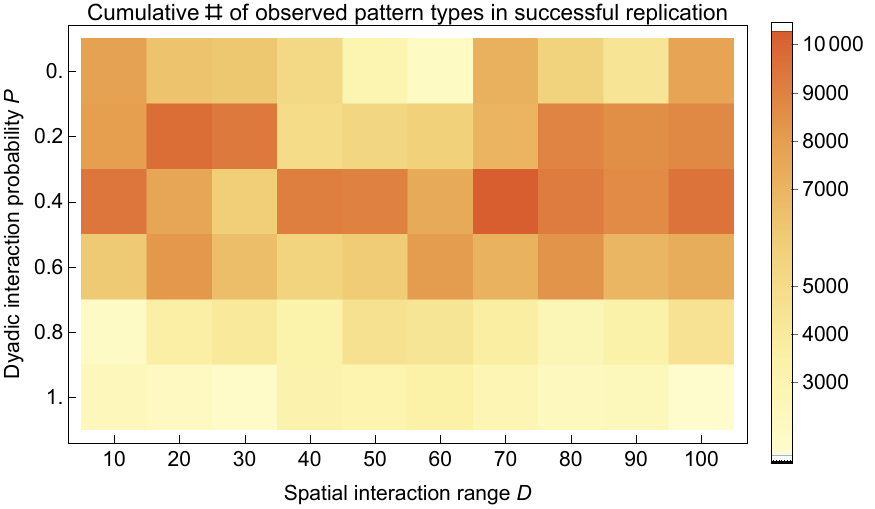} &
\includegraphics[height=1.4in]{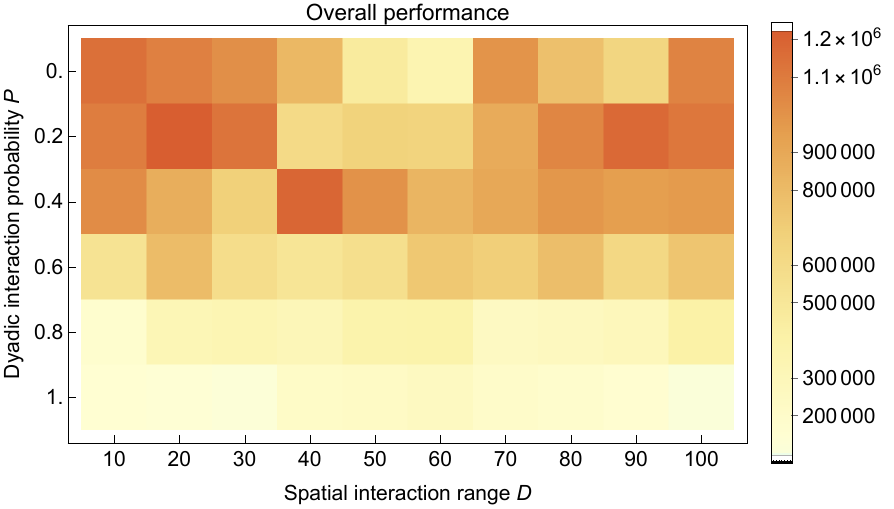}
\end{tabular}
\caption{Heatmaps showing the parameter dependence of SCHC's evolutionary dynamics on spatial interaction range $D$ and dyadic interaction probability $P$. The original setting used in \citep{sayama_schc_2025} was $(D, P) = (L, 0)$ (top right corner in each panel). {\bf (a)} Number of successful simulation runs that showed meaningful population growth (out of 30 runs). It was observed that introducing spatial locality or dyadicity significantly increased the number of successful runs. {\bf (b)} Average number of active cells involved in each successful replication event, averaged over the final 500 steps of simulation ($t = 1500-2000$). It was observed that moderate spatial locality ($D \approx 0.7 L$) helped the size growth of self-replicating patterns, especially when $P$ was small. {\bf (c)} Cumulative number of observed pattern types in successful replication at the end of simulation ($t = 2000$). It was clearly observed that the evolutionary exploration was promoted more when there was a moderate level of dyadic interaction probability, with $P \approx 0.4$ being the best. {\bf (d)} Overall performance characterized by multiplying {\bf (a)}, {\bf (b)}, and {\bf (c)}. While not a rigorous aggregation of the three metrics, this chart helps highlight the fact that the original setting $(D, P) = (L, 0)$ was not the best one for SCHC's evolutionary dynamics.}
\label{fig:P-D-final}
\end{figure}

\begin{figure}
\centering
\includegraphics[width=0.8\columnwidth]{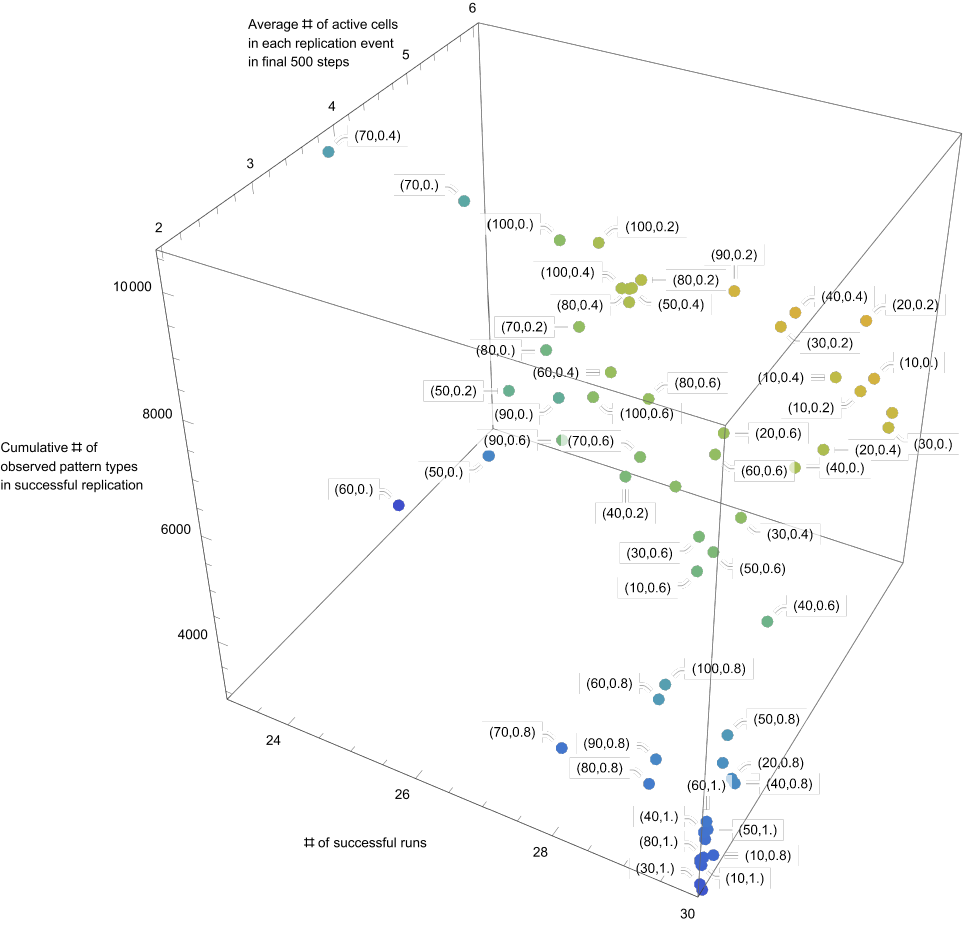}
\caption{Visualization of the simulation results showing the three outcome measurements (number of successful runs, average number of active cells in each replication event in the final 500 steps, and cumulative number of observed pattern types in successful replication) obtained under different experimental conditions ($(D, P)$, as shown in callout labels). Markers are colored according to their proximity to the ideal outcome (top right corner of the box; yellow/warm = better, blue/cold = worse). It is clearly observed that the outcome varies greatly and the conditions that resulted in better outcomes (in yellow, located near top right) tended to involve significant spatial locality and some dyadicity.}
\label{fig:3D-plot}
\end{figure}

Figures \ref{fig:P-D-final} and \ref{fig:3D-plot} show the overall summary of the three key measurements obtained from the systematic simulations we conducted (Figs.\ \ref{fig:P-D-final}a-c), together with the overall performance characterized by the product of the three measurements (Fig.\ \ref{fig:P-D-final}d) and by the 3D scatter plot (Fig.\ \ref{fig:3D-plot}). In each heatmap, the top right corner ($(D, P) = (L, 0)$) corresponds to the original setting used in the previous study \citep{sayama_schc_2025}. It was clearly observed in these plots that there were other parameter settings that produced better outcomes than the original setting. Figure \ref{fig:P-D-final}a shows that introducing spatial locality or dyadicity significantly increased the number of successful runs. Figure \ref{fig:P-D-final}b shows that reducing the spatial interaction range $D$ from $L=100$ to the smaller values promoted evolution of larger replicating patterns, especially when $P$ was small. Figure \ref{fig:P-D-final}c shows that the cumulative number of observed patterns became significantly higher when there was a moderate level of dyadic interaction probability. Finally, Figures \ref{fig:P-D-final}d and \ref{fig:3D-plot} highlighted the fact that the original setting $(D, P) = (L, 0)$ was not the best one for SCHC's evolutionary dynamics. 

It was also observed that too frequent dyadic interaction would negatively affect the evolutionary dynamics in both replicator size growth (Fig.\ \ref{fig:P-D-final}b) and evolutionary exploration (Fig.\ \ref{fig:P-D-final}c). This suggests that the evolutionary dynamics will require inherent, consistent attributes of individual entities for their competitiveness (e.g., individual-level hash values in SCHC), yet the evolutionary exploration can also be enhanced by having a moderate amount of dyadic interactions as well. Such dyadic interactions would likely have an effect to disrupt an established status quo in an evolutionary ecosystem. 

\begin{figure}
\centering
\begin{tabular}{llll}
{\bf (a)} & {\bf (b)} & {\bf (c)} & {\bf (d)}\\
\includegraphics[width=0.22\textwidth]{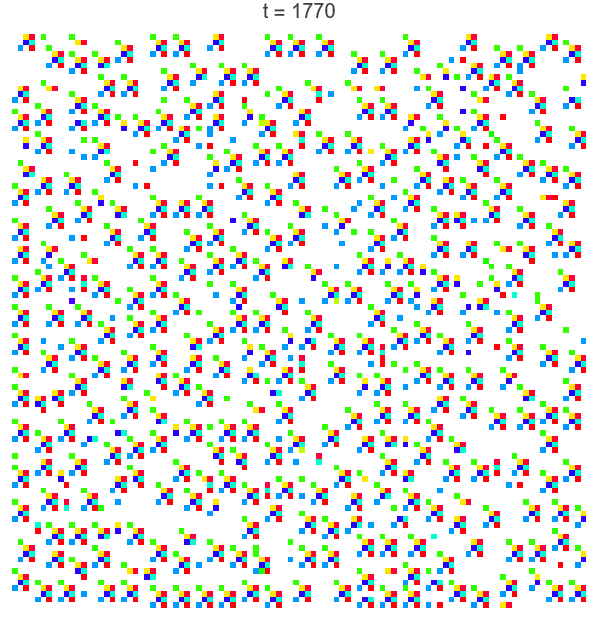} &
\includegraphics[width=0.22\textwidth]{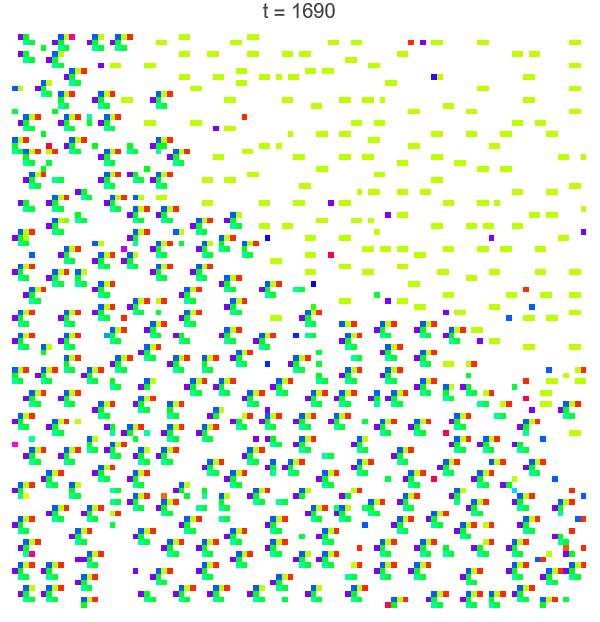} &
\includegraphics[width=0.22\textwidth]{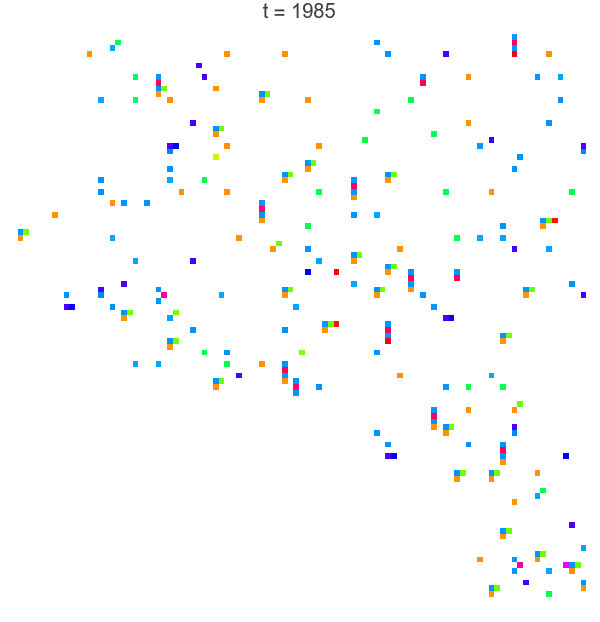} &
\includegraphics[width=0.22\textwidth]{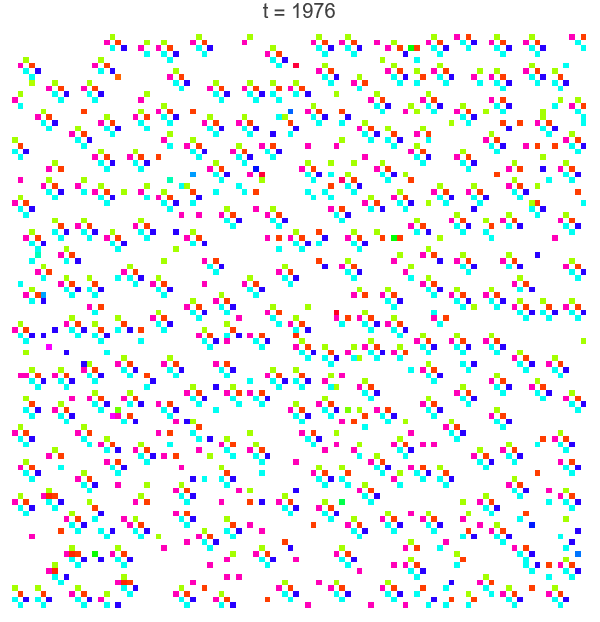} \\
\multicolumn{1}{c}{$(D, P) = (L, 0)$} &
\multicolumn{1}{c}{$(D, P) = (0.1L, 0)$} &
\multicolumn{1}{c}{$(D, P) = (L, 1)$} &
\multicolumn{1}{c}{$(D, P) = (0.2L, 0.2)$}
\end{tabular}
\caption{Exemplar configurations of simulations under four selected parameter settings. All the snapshots were taken from the later stage of the simulation.
{\bf (a)} $(D, P) = (L, 0)$, where the interactions are always global and no dyadic interactions are used (i.e., the original setting used in \citep{sayama_schc_2025}). The population tends to be homogeneous with only one self-replicating pattern dominating the entire space most of the time.
{\bf (b)} $(D, P) = (0.1L, 0)$, where the interactions are limited only to spatial neighbors. Spatially heterogeneous configurations can arise because competitions occur only locally.
{\bf (c)} $(D, P) = (L, 1)$, where the interactions are always global and dyadic. Self-replicating patterns remain simple and small, and there is hardly any dominant pattern emerging.
{\bf (d)} $(D, P) = (0.2L, 0.2)$, where the interactions have both locality and dyadicity. While the population appears to be dominated by a major self-replicating pattern, there are significantly more variations and other competing patterns present than in the original setting ({\bf (a)}).
}
\label{fig:snapshots}
\end{figure}

The effects of spatial interaction range $D$ and dyadic interaction probability $P$ were clearly recognizable in the spatio-temporal dynamics of the system's configuration when visualized and animated. Figure \ref{fig:snapshots} shows exemplar configurations of simulations under four selected parameter settings ($(D, P) = (L, 0), (0.1L, 0), (L, 1), (0.2L, 0.2)$), all taken from the later stage of the simulation\footnote{Animations of these results are available from one of the authors' YouTube channel: \url{https://www.youtube.com/ComplexSystem}}. Figure \ref{fig:snapshots}a shows a sample configuration from a simulation in the original setting, where the interactions are global and the hash value is computed for each individual pattern independently. In this setting, the population is usually dominated by only one most dominant type of self-replicating patterns with little to no variations. 
When the interactions are spatially localized (Figure \ref{fig:snapshots}b), different areas in the space can show different evolutionary dynamics, and their boundaries can propagate across space more slowly.
When the interactions are global but all dyadic (Figure \ref{fig:snapshots}c), self-replicating patterns hardly evolve to larger, more complex forms, without any dominant type emerging in the population. This is because dyadic interactions do not create an ordered linear hierarchy of dominance relationships among different patterns. 
Finally, in the optimal setting with spatial locality and dyadic interactions (Figure \ref{fig:snapshots}d), both dominance and variation can be observed simultaneously within the population. Such a balanced coexistence of exploitation and exploration enhances open-ended exploration of various pattern types, naturally including larger ones with potentially higher hash values. 

Overall, the results indicate that spatial locality and dyadicity of interactions can have significant impacts on SCHC's evolution with regard to the successful population growth, spontaneous growth of replicating pattern size, and open-ended exploration of different replicating pattern types.

\section{Extension II: Expanding the space size to observe an evolutionary transition}\label{sec:schc}

We now extend SCHC in the original setting $(D, P) = (L, 0)$ to a much larger spatial domain to explore the possibility of cross-scale emergence. This section summarizes results recently reported elsewhere \cite{horiguchi_alife2026} and adds a boundary-condition control experiment; full experimental details are in the Methods.

\subsection{A scale-controlled transition between two regimes}\label{subsec:transition}
We implemented a JAX realization of SCHC with GPU acceleration to support large ensembles and extended run lengths, preserving the published competition-replication logic but substituting a deterministic 32-bit surrogate for Mathematica's built-in \texttt{Hash}, which is not directly reproducible in JAX (Methods, Table~\ref{tab:rules}). The simulator reproduces the qualitative trends of the original study: the maximum hash score improves over time while the population-average score can peak and then decline as spatial interactions become prominent \cite{sayama_schc_2025}.

\begin{figure}
\centering
\includegraphics[width=\textwidth]{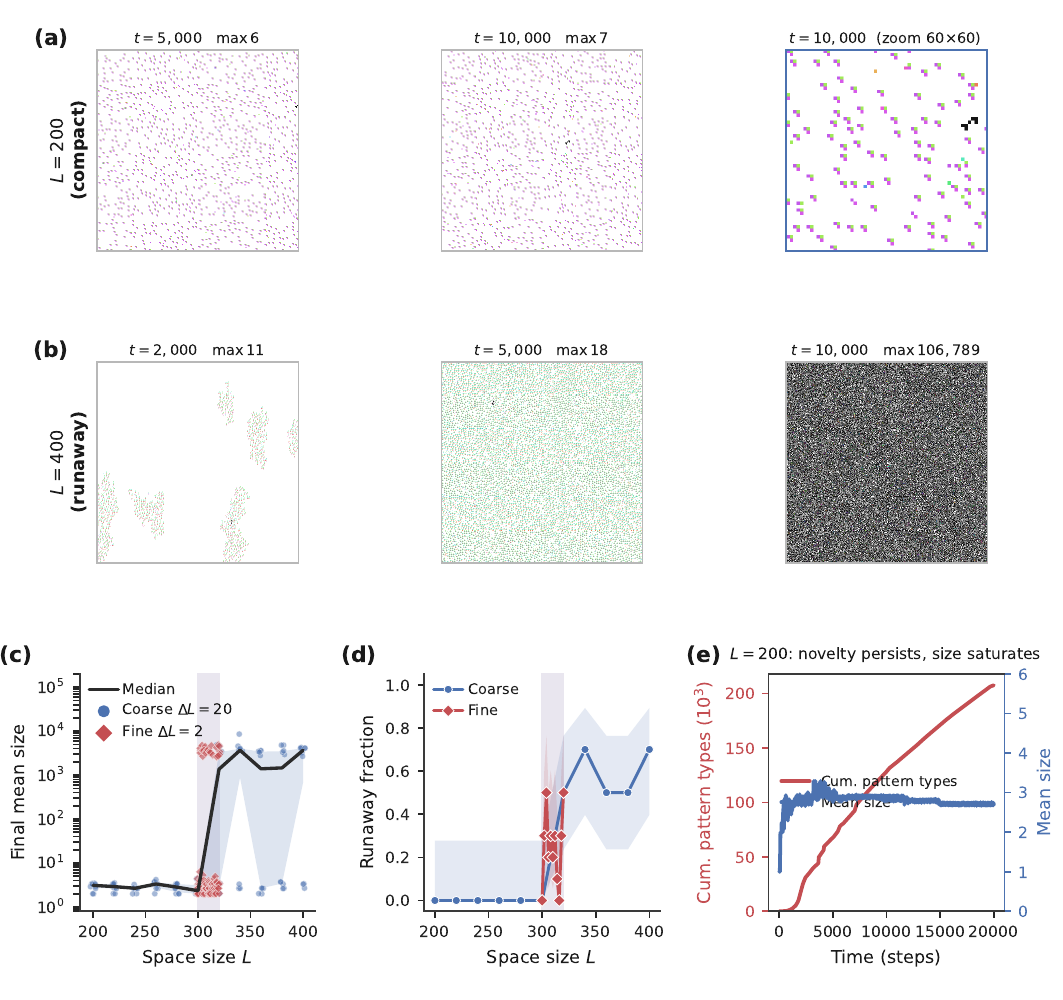}
\caption{The scale-controlled transition in SCHC. \textbf{(a)}~$L=200$ (compact regime): grid snapshots at $t=5{,}000$ and $10{,}000$ and a $60\times60$ zoom of the $t=10{,}000$ state, revealing the small self-replicating structures (largest component outlined in black; largest component $\le 7$ cells). \textbf{(b)}~$L=400$ (runaway regime): snapshots at $t=2{,}000$, $5{,}000$, $10{,}000$; distinct clusters spread and by $t=10{,}000$ the largest component reaches ${\sim}10^5$ cells and fills the space. \textbf{(c)}~Final mean component size per run versus $L$ (log scale) for the coarse ($\Delta L=20$, circles) and fine ($\Delta L=2$, squares) scans, with the ensemble median (line); grey band marks the transition window $L=300$--$320$. \textbf{(d)}~Runaway fraction versus $L$ (Wilson 95\% intervals) for the coarse and fine scans. \textbf{(e)}~Long-run novelty at $L=200$ (median over 40 runs): the cumulative number of distinct pattern types keeps rising while the mean component size saturates, so novelty persists even as a size-based complexity proxy plateaus.}\label{fig:phenomenon}
\end{figure}

Extending the simulations well beyond the horizon of the original SCHC study (100 replicates per condition at $L\in\{100,200,400\}$, 20{,}000 steps), we find that the linear size $L$ of the grid controls a transition between two qualitatively different regimes (Figure~\ref{fig:phenomenon}). In small spaces ($L\lesssim 300$) replicators stay compact (Figure~\ref{fig:phenomenon}a): typical component sizes settle at $\approx 2$--$3$ cells, and both the mean and the maximum component size show little long-run growth even over the extended horizon. Yet the cumulative number of distinct replicated patterns keeps increasing without clear saturation (Figure~\ref{fig:phenomenon}e), so sustained novelty, in the sense of newly observed spatial patterns, can persist even while a simple size-based complexity proxy plateaus; different facets of open-endedness need not coincide. In larger spaces ($L > 300$), structures instead undergo runaway growth (Figure~\ref{fig:phenomenon}b), with mean sizes reaching $\mathcal{O}(10^3)$ and maxima $\mathcal{O}(10^4$--$10^5)$ cells. A single structure comes to monopolize replication, and the hash score becomes a poor proxy for success: the temporal correlation between ensemble-mean size and ensemble-mean score turns negative across the runaway sizes (from $r\approx-0.56$ to $-0.88$; Figure~\ref{fig:mechanism}b, Table~\ref{tab:transition}), consistent with a shift from explicit score optimization to emergent competition for space.

To localize the transition we scanned $L=200$--$400$ in increments of $20$ (10 replicates per condition, 20{,}000 steps), characterizing each condition by its runaway fraction (the proportion of runs whose final mean component size exceeds $100$ cells), the late/early ratio of mean component size (steps 15{,}000--20{,}000 divided by steps 2{,}000--5{,}000), and the size--score correlation (Table~\ref{tab:transition}). A sharp but stochastic shift occurs in a narrow window between $L=300$ and $L=320$: below it mean sizes stay at $2$--$3$ cells, the runaway fraction is $0$, and the late/early ratio is $\approx 1$, whereas above it mean sizes jump to $\sim\!1{,}700$--$3{,}200$ cells, the late/early ratio exceeds unity, and the runaway fraction rises sharply (Figure~\ref{fig:phenomenon}c,d; Table~\ref{tab:transition}).

A finer scan ($L=300,302,\dots,320$, 10 replicates each) shows that the onset is not a sharp discontinuity at a single $L$ but a stochastic transition. Runaway growth already appears in $30\%$ of runs at $L=302$, and the runaway fraction fluctuates non-monotonically between $0$ and $0.5$ across $L=302$--$320$ (with no runaway at $L=300$ or $L=316$); the all-run mean final size varies likewise, from $2.6$ cells at $L=300$ to $1{,}087$ ($L=302$), $1{,}832$ ($L=304$), $815$ ($L=306$), and $1{,}855$ ($L=320$). The first-crossing times of size milestones are broad and right-skewed, and, viewed as a rate process on the larger grids ($L=320$--$400$), the fraction of runs that have nucleated grows gradually over time and faster for larger $L$ (Figure~\ref{fig:mechanism}e). These bimodal per-run outcomes are consistent with a nucleation-like mechanism in which the probability of a rare seed event grows with $L$ but remains stochastic.

The transition is not an invariant of SCHC but depends on the balance between variation and selective filtering (Figure~\ref{fig:mechanism}c,d): raising the mutation rate $\mu$ lowers the threshold (at $\mu\ge0.005$ runaway occurs even at $L=200$, eliminating the compact regime), whereas raising the per-cell death probability $p_{\mathrm{death}}$ raises or removes it, so the two act as complementary controls on the nucleation rate.

\subsection{Mechanism: a non-spatial feedback plus a finite-size effect}\label{subsec:mechanism}
The mechanism of the transition separates cleanly into a non-spatial and a spatial part. The non-spatial part is a size-biased sampling feedback: because contests are seeded by sampling cells uniformly, a component of size $s_i$ is selected with probability $1-(1-s_i/N_{\mathrm{active}})^2\approx 2s_i/N_{\mathrm{active}}$ (with $N_{\mathrm{active}}$ the total number of active cells), so larger components are contested, and therefore copied, more often. This Moran-like preferential-sampling instability is intrinsically non-spatial and can drive runaway even at $L=200$ under high mutation; it does not by itself explain the dependence on $L$. The spatial part is the finite-size medium in which this feedback operates: a growing structure must nucleate and persist within an $L\times L$ container. In this decomposition, $L$ sets the spatial budget for a seed to form and survive, rather than being a parameter of the competition rule.

\begin{figure}
\centering
\includegraphics[width=\textwidth]{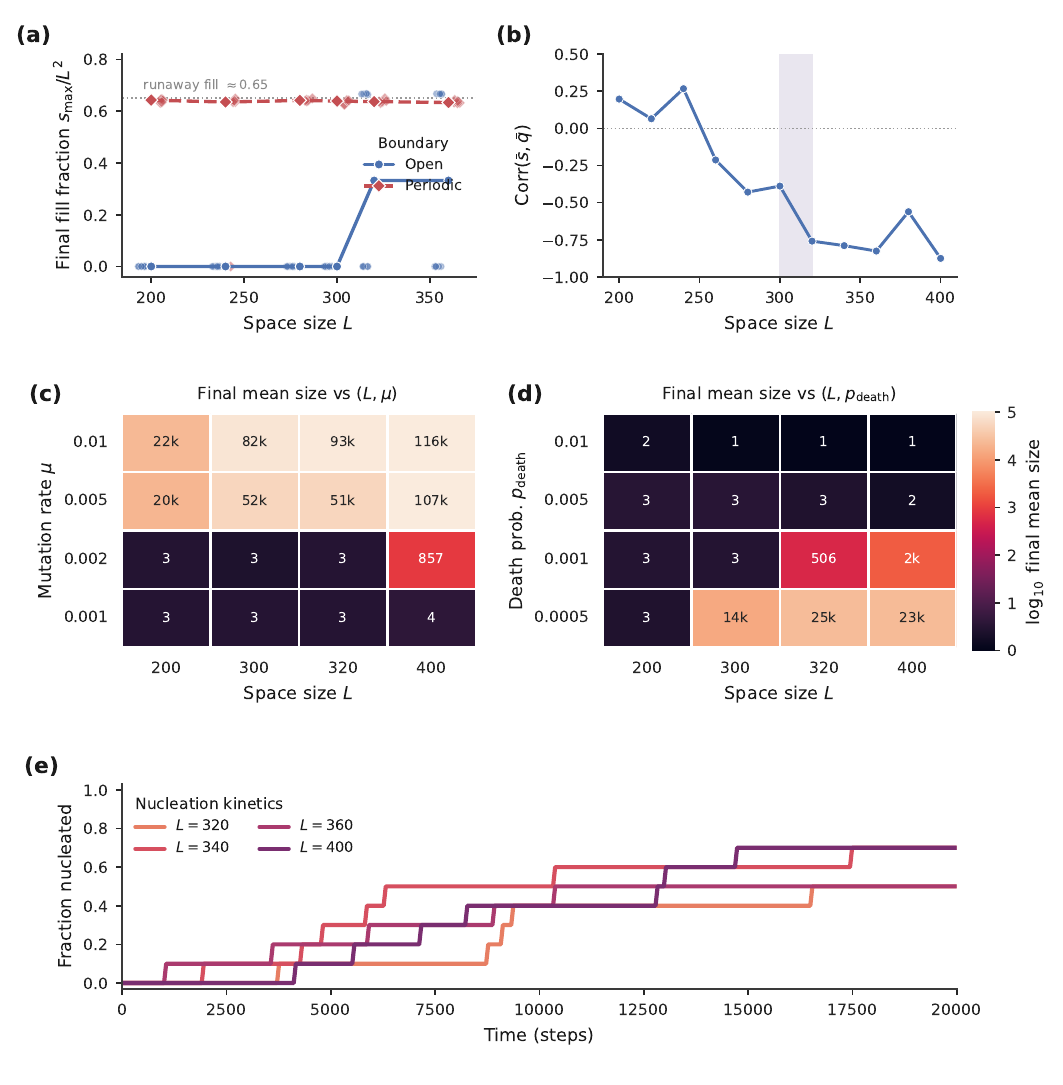}
\caption{Stochastic onset and mechanism of the transition. \textbf{(a)}~Boundary-condition control: final fill fraction $s_{\max}/L^2$ of the dominant component under open vs.\ periodic (toroidal) boundaries, otherwise identical dynamics (per-seed points, jittered in $L$, with the median line; open: 10 replicates, periodic: 5). Under open boundaries runaway is absent below $L\approx310$; under periodic boundaries it occurs at every $L$ tested. Whenever runaway occurs, the dominant structure fills a scale-invariant $\approx 0.65$ of the grid regardless of $L$ or boundary condition (dotted line), so its absolute size scales as $L^2$; the boundary sets only whether runaway happens, not how large the structure becomes. \textbf{(b)}~Temporal correlation between the ensemble-mean size and hash-score trajectories versus $L$ (grey band: transition window). \textbf{(c,d)}~Runaway phase diagrams: final mean component size (color, $\log_{10}$) over (c)~$(L,\mu)$ and (d)~$(L,p_{\mathrm{death}})$; higher mutation lowers the size threshold while higher death probability raises or removes it (annotated values are the final mean size; ``k''${=}10^3$). \textbf{(e)}~Nucleation kinetics: fraction of runs that have nucleated (mean component size $>100$) as a function of time, for $L=320$--$400$; nucleation is gradual and faster for larger $L$, as expected for a rare-seed process.}\label{fig:mechanism}
\end{figure}

A direct control experiment pins the finite-size effect on the boundary itself. We re-ran the size scan under \emph{periodic} (toroidal) boundaries, which remove edge clipping while leaving competition, scoring, and mutation unchanged, using a periodic-aware connected-component routine so that structures spanning the grid seam are correctly identified. The result is unambiguous (Figure~\ref{fig:mechanism}a): under periodic boundaries runaway occurs at every $L$ tested ($L=200$ to $360$), including every small grid ($L\le300$) at which open boundaries produce no runaway whatsoever. Nearly every periodic run enters the runaway regime (29 of 30 runs, runaway fraction $\approx 0.97$; the sole exception is a single run at $L=240$), versus a runaway fraction of $0$ under open boundaries below the transition. Removing the boundary therefore removes the finite-size threshold entirely.

Figure~\ref{fig:mechanism}a measures the outcome as the footprint of the dominant component, the fill fraction $s_{\max}/L^2$, rather than the mean component size. So measured, the picture is clean: whenever a structure runs away, under periodic boundaries at every $L$, and under open boundaries above the transition, it converges to a scale-invariant fill fraction of $\approx 0.65$, essentially independent of both $L$ and the boundary condition. Its \emph{absolute} size therefore grows in proportion to the available area ($s_{\max}\propto L^2$; under periodic boundaries the dominant component contains $\approx 2.6\times10^{4}$ cells at $L=200$ and $\approx 8.2\times10^{4}$ at $L=360$), set by a dynamic balance between local replicative growth and per-cell death rather than by the grid size. The boundary condition governs only \emph{whether} a structure runs away, not \emph{how large} it becomes once it does. (An ensemble mean over all components would conflate this scale-invariant giant with a number of small debris clusters that itself grows with the area, and so appears nearly flat in $L$; the fill fraction of the dominant component is the informative quantity.)

This confirms the decomposition directly: the size-biased sampling feedback drives runaway intrinsically and independently of $L$, while the open boundary is precisely what gates the phenomenon in small grids, clipping a growing structure's footprint before it can grow large enough to monopolize replication. The linear scale $L$ controls the transition only through the open boundary; it is not a parameter of the competition rule.

This mechanism also accounts for the isolated explosive-growth event flagged in the spatial-dyadic extension at $(D,P)=(10,0.6)$ (Section~\ref{sec:dyadic}): strong local, dyadic competition can occasionally seed a runaway structure even in the small ($L=100$) grid used there, much as elevated mutation induces runaway at small $L$ under global competition (Figure~\ref{fig:mechanism}c). Such events are rare, and were therefore excluded from the Extension~I statistics, but they reflect the same nucleation-like runaway analyzed here rather than a distinct anomaly.

\begin{table}[t]
\caption{Summary of the SCHC transition scan ($L=200$--$400$, 10 replicates per condition, 20{,}000 steps). Values are means across replicates unless otherwise noted. The narrow jump between $L=300$ and $L=320$ marks the stochastic transition from the score-optimization regime to the size-dominance regime.}\label{tab:transition}
\begin{tabular*}{\textwidth}{@{\extracolsep\fill}rrrrrr}
\toprule
$L$ & Mean size & Max size & Late/Early & Runaway frac. & $r(\bar{s}, \bar{q})$ \\
\midrule
200 & 2.9 & 6.8 & 0.99 & 0.0 & 0.20 \\
220 & 2.9 & 9.6 & 0.96 & 0.0 & 0.07 \\
240 & 2.6 & 6.6 & 0.99 & 0.0 & 0.27 \\
260 & 3.2 & 15.6 & 1.01 & 0.0 & $-$0.21 \\
280 & 2.8 & 12.4 & 0.92 & 0.0 & $-$0.43 \\
300 & 2.6 & 13.9 & 1.06 & 0.0 & $-$0.39 \\
\midrule
320 & 1{,}855 & 34{,}111 & 12.1 & 0.5 & $-$0.76 \\
340 & 3{,}190 & 53{,}997 & 6.19 & 0.7 & $-$0.79 \\
360 & 1{,}678 & 43{,}176 & 3.63 & 0.5 & $-$0.83 \\
380 & 1{,}941 & 48{,}053 & 2.80 & 0.5 & $-$0.56 \\
400 & 2{,}657 & 74{,}601 & 35.8 & 0.7 & $-$0.88 \\
\botrule
\end{tabular*}
\footnotetext{``Mean size'' and ``Max size'' are measured at step 20{,}000. ``Late/Early'' is the ratio of mean component size in the late window (steps 15{,}000--20{,}000) to the early window (steps 2{,}000--5{,}000). ``Runaway frac.'' is the fraction of runs with final mean component size $> 100$. $r(\bar{s}, \bar{q})$ is the Pearson correlation between the ensemble-mean size trajectory and the ensemble-mean hash-score trajectory over steps 2{,}000--10{,}000.}
\end{table}

\section{Discussion}\label{sec:discussion}

We extended SCHC in two different ways to explore its evolutionary dynamics. The first extension with spatial locality and dyadic nature of competitive interactions better captures more nuanced ecological interactions among replicating patterns without requiring substantial additional computational costs. Numerical experiments revealed that spatial locality and dyadicity each had a positive effect on successful population growth, pattern size growth, and open-ended exploration in the evolution dynamics of SCHC. More specifically, the spatial locality promoted replication, population growth, and size growth of self-replicating patterns, whereas the moderate level of dyadic interactions significantly increased the number of explored self-replicating pattern types during the evolution. We were able to identify some sweet spots in the parameter space where the evolutionary dynamics were significantly  enhanced compared to the original setting used in previous studies.

The result re-emphasizes a well-established insight that the nature of ecological interactions among replicating entities has a significant impact on the dynamics of their evolution over time and space. Both locality and dyadicity have positive effects on the diversity of self-replicating patterns in the course of evolution. Spatial locality naturally separates one spatial region from another, creating geographical diversities of the evolutionary processes that differ from place to place. Meanwhile, dyadic interaction allows for unordered, non-hierarchical networks of ecological dominance relationships, including circular ones like those in the Rock-Paper-Scissors game. Such unordered ecological relationships make it impossible for any single replicating pattern to dominate the entire population, making the evolutionary process more dynamic and more exploratory. Maintaining the ``right balance'' between exploration and exploitation can lead to a greater chance of successful population growth and evolutionary discoveries of diverse forms, including more complex ones.

The second extension identifies a scale-controlled reorganization of evolution: small spaces favor compact structures with weak size-score coupling, whereas large spaces permit a size-dominance regime in which sufficiently large structures monopolize replication and decouple success from the hash score. This clarifies a point central to the whole Hash Chemistry programme, that the explicit hash score is not evolutionary fitness; realized fitness emerges from interaction, interference, and the ability to persist and replicate in the medium.

The shift is not a strict thermodynamic phase transition. Statistical testing confirms a significant jump in mean component size between $L=300$ and $L=320$ (Welch's $t=-2.85$, $p=0.019$; Mann--Whitney $U=18$, $p=0.017$; Cohen's $d=-1.27$), yet the fine scan shows runaway growth occurring stochastically at multiple $L$, with non-monotonic runaway fraction and broad waiting-time distributions, pointing to rare seed formation rather than a clean discontinuity at a single critical size. Its decomposition into a non-spatial size-biased sampling feedback and a finite-size effect, together with the boundary-control experiment, shows that a purely quantitative change in scale can flip the system into a qualitatively different collective regime, and that the responsible ingredients can be individually identified in a minimal model.

More broadly, the size-dependent advantage seen here, in which larger structures gain disproportionate access to replication simply by occupying more space, offers a minimal analogue of resource monopolization in evolution. The major transitions from replicating molecules to cells and from unicellular to multicellular organisms involved not only increases in intrinsic fitness but also the capacity to commandeer a larger share of physical and energetic resources \cite{maynardsmith1997}. Whether spatial monopolization of the kind seen in SCHC captures any essential feature of such transitions remains open, but the model provides a concrete, computationally tractable setting in which to study it \cite{adami2000}. Rich adaptive dynamics often arise near the boundary between collective regimes \cite{langton1990,kauffman1993}, and the Hash Chemistry family offers minimal, mechanistically transparent instances in which such boundaries can be dissected. We note that truly local competition (as opposed to the global competition with local replication studied here) is exactly what the spatial-dyadic extension of SCHC introduces through its interaction-range parameter (Section~\ref{sec:dyadic}) \cite{sayama_spatialdyadic_2026}; the present study deliberately isolates the global-competition regime.

Several limitations bound these conclusions. The finite-size transition is characterized from modest ensembles, ten runs per grid in the size scans and five per condition in the sensitivity and boundary-control experiments, so the precise location and width of the transition window, and the fine-scan runaway fractions, carry appreciable statistical uncertainty. The periodic-boundary control additionally used a shorter run horizon ($5{,}000$ versus $20{,}000$ steps); this is ample for the rapid nucleation seen under toroidal boundaries but leaves the long-time periodic behavior less thoroughly sampled than the open-boundary case. Our GPU implementation also substitutes a deterministic 32-bit surrogate for Mathematica's built-in \texttt{Hash} (Methods); we verified qualitative agreement with the original dynamics but not an exact match, so results that hinge on the detailed hash statistics should be read with that caveat. Finally, the size-based complexity proxy is a deliberately simple summary rather than a rigorous measure of structural complexity. None of these caveats undermines the central, boundary-controlled decomposition, which holds consistently across the open and periodic conditions, but each marks a direction in which a larger or more carefully instrumented study would sharpen the picture.

\section{Conclusion}\label{sec:conclusion}
Hash Chemistry provides a family of minimal evolutionary models in which a deterministic hash opens a combinatorially vast possibility space while keeping the dynamics tractable. Reviewing the family alongside two new SCHC extensions, we have shown how the same core idea supports the study of open-ended novelty. In particular, with SCHC extensions, we showed that spatial locality and dyadicity of interactions enrich its evolutionary dynamics and that the size of the space controls a shift from compact replicators to a runaway size-dominance regime, with a mechanism that separates into a non-spatial size-biased feedback and a finite-size spatial effect. Altogether, these results illustrate the rich potential of Hash Chemistry as a minimal, mechanistically transparent testbed for studying open-ended evolution across scales.

There are still many possibilities and directions of further model extensions, including (but not limited to) introducing spatially and/or temporally heterogeneous hash functions, expanding the spatial structure from 2D to 3D or higher dimensions or even to more complex network topologies, and even allowing self-replicating patterns to dynamically modify the underlying local network topology using adaptive network approaches \citep{gross2009adaptive,sayama2013modeling}\footnote{We see this direction particularly intriguing because the current formulation of SCHC is technically no more than a special case of Generative Network Automata \citep{sayama2007generative,sayama2009generative,sayama2013modeling} based on repeated extraction and replacement of labeled subgraphs on a network (e.g., a 2D grid in SCHC).}. Another promising direction of further research is to explore other open-ended evaluation mechanisms than generic hash functions, such as large language models (LLMs) \citep{suzuki2025evolutionary,suzuki2025b} and other modern AI technologies that allow for open-ended evaluations of free-form, higher-order, unboundedly complex self-replicating structures. We have been using hash functions as a simple, fast, ready-to-use surrogate of the (fundamentally unknowable) ``oracle'' universal evaluation mechanisms, but hash functions themselves are not an essential part of this line of research. The key idea has always been the {\em cardinality leap}, the possibility for any higher-order entities made of smaller-scale replicating units to be able to participate in and be evaluated through the evolutionary process \citep{sayama_hashchem_2019}. Any computational or physical mechanisms that allow for it can be actively explored to achieve truly open-ended artificial evolutionary systems.

\section{Methods}\label{sec:methods}

\subsection{Model description}\label{subsec:model}
Structural Cellular Hash Chemistry (SCHC) \cite{sayama_schc_2025} operates on an $L \times L$ 2D grid in which each cell is either empty (state 0) or carries a discrete type from $\{1, \dots, k\}$. Individuals are connected components of non-empty cells under 8-connectivity (Moore neighborhood). In the original version and the first extension described above, we used Mathematica's built-in \texttt{Hash} function. In the second extension for which we developed a JAX realization, we replaced Mathematica's \texttt{Hash} with a deterministic 32-bit surrogate score because the former is not directly reproducible in JAX. Each step of simulation consists of one or more pairwise competitions that advance the system clock by one logical time unit (Algorithm~\ref{alg:schc}); Table~\ref{tab:rules} lists the rule-level implementation choices.

\subsection{Competition and replication}\label{subsec:competition}
Each competition proceeds as follows:
\begin{enumerate}
\item Two active cells are sampled uniformly at random with replacement from the set of non-empty cells.
\item Their connected components $C_1$ and $C_2$ are identified via flood-fill using 8-connectivity.
\item Each component is canonicalized by translating its coordinates so the minimum $x$- and $y$-coordinates are zero.
\item A hash-based score $s(C_i)$ is computed (using Mathematica's \texttt{Hash}, normalized to $[0, 1]$, except for the second extension where Equation~\ref{eq:score} was used instead).
\item The higher-scoring component wins deterministically; exact ties are broken uniformly at random. The winner's padded bounding box is centered on the loser's bounding-box center and copied with per-cell death and mutation (Equation~\ref{eq:mutation}).
\end{enumerate}

Sampling is at the cell level rather than the component level, so competition is global (well-mixed) but larger components participate more often by occupying more cells. If both sampled cells belong to the same component, the rewrite acts as a self-copy with mutation. The grid uses open boundaries: any part of the copied block outside the $L\times L$ domain is clipped. A periodic-boundary variant, used for the control experiment in Section~\ref{subsec:mechanism}, instead wraps the copied block toroidally and identifies connected components with a periodic-aware routine. After each competition the clock advances by $p_w = |C_w|/N_{\mathrm{active}}$, where $|C_w|$ is the winner's size and $N_{\mathrm{active}}$ the total number of non-empty cells; a step completes when the accumulated clock exceeds~1, so one step corresponds roughly to one ``generation.''

For the first extension with spatial locality and dyadicity of interactions, the algorithm was revised as follows, where the two additional parameters were used in the revised parts (highlighted in bold fonts):
\begin{enumerate}
\item {\bf One} active cell is sampled uniformly at random from the set of non-empty cells.
\item {\bf Another active cell is sampled uniformly at random from the set of non-empty cells within chessboard distance $D$ of the first chosen cell.} 
\item Their connected components $C_1$ and $C_2$ are identified via flood-fill using 8-connectivity.
\item Each component is canonicalized by translating its coordinates so the minimum $x$- and $y$-coordinates are zero.
\item {\bf Do one of the following:}
\begin{enumerate}
\item {\bf With probability $1-P$,} hash-based scores $s(C_i)$ are computed independently for the two components, and the winning pattern is determined the same way as before. {\em (individual competition)}
\item {\bf With probability $P$, the two components are combined in a sorted order and a hash value of the entire combined list is calculated. If the hash value is greater than 0.5, the first pattern is chosen as the winner; otherwise the second one is chosen as the winner.} {\em (dyadic interaction)}
\end{enumerate}
\item The winner's padded bounding box is centered on the loser's bounding-box center and copied with per-cell death and mutation.
\end{enumerate}
In the above Step 5(b), the hash value evaluation of a combined list represents direct, {\em dyadic} interaction between two replicating patterns, because it makes the chance of successful replication dependent on who the competitor is, not just decided by the inherent individual-level hash values. Such dyadic interactions can create a nontrivial network of unordered, potentially circular, ecological dominance relationships. This modeling approach is similar to and inspired by the recent work by \cite{suzuki2025evolutionary,suzuki2025b}, where they evolved words by submitting queries to LLMs about which of the two words would be stronger. In our SCHC model, hash functions play similar roles of a generic ``oracle'' evaluator as LLMs in their work.

\subsection{Hash-based score function used in JAX implementation}\label{subsec:fitness}
The score for a component $C$ with canonicalized cell positions $\{(x_i, y_i)\}$ and cell types $\{v_i\}$ is a deterministic 32-bit surrogate hash:
\begin{equation}\label{eq:score}
s(C) = \frac{\left(\left[\left(\bigoplus_{i \in C} h(x_i, y_i, v_i)\right) \oplus c_3\right] \cdot c_4\right) \bmod 2^{32}}{2^{32} - 1},
\end{equation}
where $\oplus$ denotes bitwise XOR and each cell's contribution is
\begin{equation}
h(x, y, v) = (x \cdot c_1) \oplus (y \cdot c_2) \oplus v,
\end{equation}
with constants $c_1 = \texttt{0x9E3779B9}$, $c_2 = \texttt{0x85EBCA6B}$, $c_3 = \texttt{0x45D9F3B}$, and $c_4 = \texttt{0x27D4EB2D}$ in 32-bit unsigned arithmetic. The result is normalized to $[0, 1]$. We call this a hash score rather than fitness because realized replication success is emergent and depends on spatial interactions.

\subsection{Mutation}\label{subsec:mutation}
When the winner is copied onto the loser's region, each cell undergoes independent stochastic perturbation:
\begin{equation}\label{eq:mutation}
v'_i = \begin{cases}
0 & \text{with probability } p_{\mathrm{death}}, \\
\text{Uniform}\{1, \dots, k\} & \text{with probability } (1 - p_{\mathrm{death}}) \cdot \mu, \\
v_i & \text{otherwise},
\end{cases}
\end{equation}
where $p_{\mathrm{death}}$ is the cell death probability and $\mu$ the conditional mutation probability given survival. With the defaults $p_{\mathrm{death}} = 0.001$ and $\mu = 0.002/0.999$ the effective per-cell mutation probability is exactly $0.002$. Table~\ref{tab:params} lists the default parameter values used throughout.

\begin{table}[t]
\caption{Rule-level implementation choices in the JAX realization.}\label{tab:rules}
\begin{tabular}{@{}lp{0.72\textwidth}@{}}
\toprule
Aspect & JAX implementation \\
\midrule
Score function & Deterministic 32-bit surrogate score on canonicalized relative coordinates and cell types; used because Mathematica's built-in \texttt{Hash} is not directly reproducible in JAX. \\
Boundary condition & Open boundaries; copied cells outside the $L \times L$ domain are clipped. (A periodic/toroidal variant is used for the boundary-control experiment in Section~\ref{subsec:mechanism}.) \\
Component detection & Flood-fill with 8-connectivity (Moore neighborhood); periodic-aware under toroidal boundaries. \\
Competition sampling & Two active cells sampled uniformly at random with replacement from all non-empty cells (global/well-mixed). \\
Winner selection & Higher-scoring component wins deterministically; exact ties broken uniformly at random. \\
Rewrite operator & Winner's padded bounding box (one-cell margin) centered on the loser's bounding-box center and copied with per-cell death and mutation. \\
Same-component sampling & Allowed; acts as self-copy with mutation. \\
Clock increment & Each contest advances time by $|C_w|/N_{\mathrm{active}}$; a step accumulates contests until elapsed time reaches 1. \\
\botrule
\end{tabular}
\end{table}

\begin{table}[t]
\caption{SCHC simulation parameters.}\label{tab:params}
\begin{tabular}{@{}llp{22em}@{}}
\toprule
Parameter & Value & Description \\
\midrule
$k$ & 1{,}000 & Number of possible cell types \\
$n$ & 10 & Initial number of randomly placed active cells \\
$L$ & varied & Linear grid dimension ($L \times L$ grid) \\
$\mu$ & $0.002/0.999$ & Conditional mutation probability (chosen so the effective rate is 0.002) \\
$p_{\mathrm{death}}$ & 0.001 & Per-cell death probability during copying \\
\botrule
\end{tabular}
\end{table}

\subsection{Experimental design in Extension II}\label{subsec:experiments}
\paragraph{Long-run comparison.} $L\in\{100,200,400\}$, 20{,}000 steps, 100 GPU-parallelized replicates per condition. Seeds were consecutive integers from 0.

\paragraph{Coarse transition scan.} $L=200,220,\dots,400$ (increments of 20 in grid side length), 20{,}000 steps, 10 replicates per condition. We recorded mean/maximum component size, mean/maximum hash score, and cumulative distinct-pattern counts at regular intervals, and computed the size-score correlation over steps 2{,}000--10{,}000.

\paragraph{Fine transition scan.} $L=300,302,\dots,320$ (increments of 2), 20{,}000 steps, 10 replicates per condition; metrics sampled every 10 steps.

\paragraph{Mutation-rate sensitivity.} $\mu\in\{0.001,0.002,0.005,0.01\}$ crossed with $L\in\{200,300,320,400\}$, 10{,}000 steps, 5 replicates per condition.

\paragraph{Death-probability sensitivity.} $p_{\mathrm{death}}\in\{0.0005,0.001,0.005,0.01\}$ crossed with $L\in\{200,300,320,400\}$, 10{,}000 steps, 5 replicates per condition.

\paragraph{Boundary-condition control.} The size scan repeated under periodic (toroidal) boundaries at $L\in\{200,240,280,300,320,360\}$ (5 replicates per condition), matching the open-boundary scan (10 replicates) in all other respects except the run horizon, to isolate the role of the boundary. Periodic runs used a horizon of up to $5{,}000$ steps (rather than $20{,}000$); under periodic boundaries runaway, when it occurs, nucleates well within this budget. Connected components are identified with a periodic-aware routine (wrapped/rolled neighborhood dilation) so that structures spanning the grid seam are correctly counted; both the replication rewrite and the component detection are fully toroidal. Runs were truncated once the mean component size exceeded a large threshold, at which point runaway is unambiguous, to cap the heavy flood-fill cost on grid-spanning components.

\subsection{Implementation}\label{subsec:implementation}
The codes for numerical simulation, data analysis and visualization for earlier Hash Chemistry models and the original/spatial-dyadic SCHC were implemented in Wolfram Mathematica 14.3. The computational hardware used for these numerical simulations was the same as used in \citep{sayama_schc_2025}. 

The simulator used in the second extension of the present study was implemented in JAX \cite{jax2018}. Core operations (flood-fill connected-component identification, hash-score computation, and mutation) are JIT-compiled device-side operations using \texttt{jax.lax.while\_loop} and \texttt{jax.lax.fori\_loop}, executing entirely on GPU without host-device transfers; the open-boundary connected-component routine uses \texttt{jax.lax.reduce\_window} with a $3\times3$ max-pooling kernel over Boolean occupancy masks, and the periodic variant uses wrapped (rolled) shifts. Processed per-seed and summary CSV files used for the figures are included in the repository under \texttt{results/}.

\begin{algorithm}[t]
\caption{One SCHC simulation step}\label{alg:schc}
\begin{algorithmic}[1]
\Require Grid $G$ of size $L \times L$, parameters $k$, $\mu$, $p_{\mathrm{death}}$
\State $\text{elapsed} \gets 0$
\While{$\text{elapsed} < 1$}
    \State Select two random active cells $c_1, c_2$ from $G$
    \State $C_1 \gets \Call{FloodFill}{G, c_1}$ \Comment{8-connected component}
    \State $C_2 \gets \Call{FloodFill}{G, c_2}$
    \State $s_1 \gets \Call{HashScore}{C_1}$; \quad $s_2 \gets \Call{HashScore}{C_2}$
    \State $\text{winner}, \text{loser} \gets \Call{HigherScoreSelect}{C_1, C_2, s_1, s_2}$
    \State Copy the winner's padded bounding box onto the loser's centered region with per-cell death and mutation
    \State $\text{elapsed} \gets \text{elapsed} + |\text{winner}| / N_{\mathrm{active}}$
\EndWhile
\end{algorithmic}
\end{algorithm}

\backmatter

\bmhead{Acknowledgements}
Computational resources were provided by Binghamton University and the University of Tokyo. I.H.\ used Claude Code as an AI-assisted tool during the preparation of this work, including for code implementation, figure-generation support, manuscript editing, and reference checking; all scientific claims, analyses, and interpretations were verified and approved by the authors, who take full responsibility for the content.

\section*{Declarations}

\begin{itemize}
\item \textbf{Funding:} Not applicable.
\item \textbf{Competing interests:} The authors declare no competing interests.
\item \textbf{Prior publication:} Shorter studies of the spatial-dyadic SCHC (Section~\ref{sec:dyadic}) and the SCHC finite-size transition (Section~\ref{sec:schc}) were presented at the Artificial Life Conference 2026 \cite{sayama_spatialdyadic_2026,horiguchi_alife2026}. The present paper is a broader review of the Hash Chemistry family of models that incorporates a condensed account of that result together with new material.
\item \textbf{Data availability:} Processed per-seed and summary CSV files from the second extension are included in the repository under \texttt{results/}; additional raw outputs are available from the corresponding author on reasonable request.
\item \textbf{Code availability:} The original SCHC, the spatial-dyadic SCHC, and the JAX implementation of SCHC are available at
\url{https://github.com/hsayama/Structural-Cellular-Hash-Chemistry},
\url{https://github.com/hsayama/Spatial-Dyadic-Structural-Cellular-Hash-Chemistry}, and
\url{https://github.com/NeoGendaijin/py-hash-chemistry}, respectively.
\item \textbf{Author contribution:} I.H.\ implemented the JAX simulator, designed and ran the second extension SCHC experiments, and analyzed the data. H.S.\ conceived the Hash Chemistry family of models, implemented the earlier model variants and spatial-dyadic SCHC, designed and ran the first extension SCHC experiments, analyzed the data, and supervised the overall project. Both authors wrote, revised, reviewed and approved the manuscript.
\end{itemize}

\bibliography{horiguchi-sayama-2026}

\end{document}